\documentclass[sigconf, nonacm]{acmart}
\usepackage{algorithm}
\usepackage{algorithmic}
\usepackage{textcomp}
\usepackage{xspace}
\usepackage{color,soul}
\usepackage{amsbsy}
\usepackage{pifont}
\usepackage{listings}
\newcommand{\proj}{SVApshot\xspace}

\definecolor{dred}{rgb}{0.8, 0.0, 0.0}
\definecolor{dgreen}{rgb}{0.0, 0.5, 0.0}
\definecolor{dyel}{rgb}{1.0, 0.77, 0.05}
\AtBeginDocument{%
  \providecommand\BibTeX{{%
    \normalfont B\kern-0.5em{\scshape i\kern-0.25em b}\kern-0.8em\TeX}}}
\setcopyright{cc}
\setcctype{by}
\copyrightyear{2026}
\graphicspath{{./images/}} 
\begin{document}
\title[]{Will AI mess my RTL? A snapshot-based approach to verifying AI-written RTL}
\thanks{Paper presented at the DAC 2026 Work in Progress (WIP) Poster Session.}
\author{Bernat Homs}
\affiliation{%
  \institution{Universitat Polit\`ecnica de Catalunya}
  \institution{Barcelona Supercomputing Center}
  \city{Barcelona}
  \country{Spain}}
\email{bernat.homsgisbert@bsc.es}
\author{Oscar Palomar}
\affiliation{%
  \institution{Barcelona Supercomputing Center}
  \city{Barcelona}
  \country{Spain}}
\email{oscar.palomar@bsc.es}
\author{Miquel Moreto}
\affiliation{%
  \institution{Universitat Polit\`ecnica de Catalunya}
  \institution{Barcelona Supercomputing Center}
  \city{Barcelona}
  \country{Spain}}
\email{miquel.moreto@bsc.es}
\author{Marcelo Orenes-Vera}
\affiliation{%
  \institution{NVIDIA}
  \city{Santa Clara}
  \country{USA}}
\email{morenesvera@nvidia.com}
\renewcommand{\shortauthors}{Homs et al.}
\makeatletter
\def\@mkauthors{%
  \global\setbox\mktitle@bx=\vbox{\noindent\unvbox\mktitle@bx\par\medskip
    \noindent
    \begin{minipage}[t]{0.25\textwidth}\centering
      {\Large\sffamily Bernat Homs\textsuperscript{1,2}}\par
      {\normalsize bernat.homsgisbert@bsc.es}
    \end{minipage}%
    \begin{minipage}[t]{0.25\textwidth}\centering
      {\Large\sffamily Oscar Palomar\textsuperscript{2}}\par
      {\normalsize oscar.palomar@bsc.es}
    \end{minipage}%
    \begin{minipage}[t]{0.25\textwidth}\centering
      {\Large\sffamily Miquel Moreto\textsuperscript{1,2}}\par
      {\normalsize miquel.moreto@bsc.es}
    \end{minipage}%
    \begin{minipage}[t]{0.25\textwidth}\centering
      {\Large\sffamily Marcelo Orenes-Vera\textsuperscript{3}}\par
      {\normalsize morenesvera@nvidia.com}
    \end{minipage}\par
    \medskip
    {\normalsize\normalfont\centering
      \textsuperscript{1}Universitat Polit\`ecnica de Catalunya, Barcelona, Spain\par
      \textsuperscript{2}Barcelona Supercomputing Center, Barcelona, Spain\par
      \textsuperscript{3}NVIDIA, Santa Clara, USA\par}
    \vspace{6pt}%
  }%
}
\makeatother
\begin{abstract}
With the growing interest in using AI (Artificial Intelligence) for RTL (Register-Transfer Level) hardware development, robust and comprehensive verification has become more important than ever.
As Large Language Models (LLMs) increasingly assist in creating and modifying RTL designs, ensuring that these changes preserve existing functionality is paramount for building trust in AI-driven workflows.
We present SVApshot, a fully automated framework that leverages LLMs to generate and systematically attempt to correct SystemVerilog Assertions (SVA) for RTL modules, completely removing the human from the generation loop.
SVApshot introduces a novel snapshot methodology that captures the current functionality of a design as a comprehensive set of formal assertions, creating a regression suite for validating future changes---whether made by humans or AI.
The framework features automated duplicate detection, assertion set expansion, and iterative repair of failing assertions using LLM-guided debugging with formal verification feedback.
Our experimental evaluation across diverse RTL  modules demonstrates high coverage for complex modules, with successful detection of both manually-injected and AI-introduced bugs.
Additionally, this framework opens the door for automatic generation of formal testbenches to benchmark LLMs for hardware design, moving beyond binary pass/fail metrics towards nuanced assertion-based scoring.
\end{abstract}
\maketitle
\vspace{-1.5mm}
\section{Introduction and Background}\label{sec:intro}
The rapid advancement of LLMs has opened a new era of AI-driven RTL design, where automated tools increasingly handle the creation and modification of hardware descriptions~\cite{rtlcoder2024,aivril2024,rtlpp2025}, introducing unprecedented verification challenges.
As LLMs automate complex design tasks, ensuring that AI modifications to RTL preserve existing functionality while introducing new features has become paramount for building trust in AI-driven workflows.
Traditional approaches to automated assertion generation have made significant strides but fall short of addressing the unique demands of AI-assisted hardware design. Early methods like GoldMine~\cite{goldmine2013} pioneered dynamic assertion mining through simulation-based data mining, while template-based approaches such as HARM~\cite{harm2022} and A-TEAM~\cite{Danese2017ATeam} leveraged predefined patterns to generate assertions from execution traces.
ILA~\cite{ilang} generates a model of the design from a functional specification and compares it against its RTL implementation, and AutoSVA~\cite{autosva} introduced annotation-driven formal testbench generation. 
However, these conventional methods are constrained by their dependence on simulation coverage, predefined templates, models, or manual annotations.
The emergence of LLM-based assertion generation has shown promise in addressing some of these limitations~\cite{autosva2,assertllm2025}.
Recent works like AssertLLM~\cite{assertllm2025} process complete specification documents to generate assertions, while AssertionForge~\cite{assertionforge2025} constructs knowledge graphs from both specifications and RTL code to discover complex signal interactions.
Spec2Assertion~\cite{Menon2025Spec2Assertion} focuses on pre-RTL assertion generation from architectural specifications.
Despite these advances, current LLM-based approaches suffer from critical gaps: they lack mechanisms for iterative refinement of failing assertions and do not provide systematic approaches for maintaining assertions across design evolution.
\begin{figure}[t]
\centering
\includegraphics[width=\columnwidth]{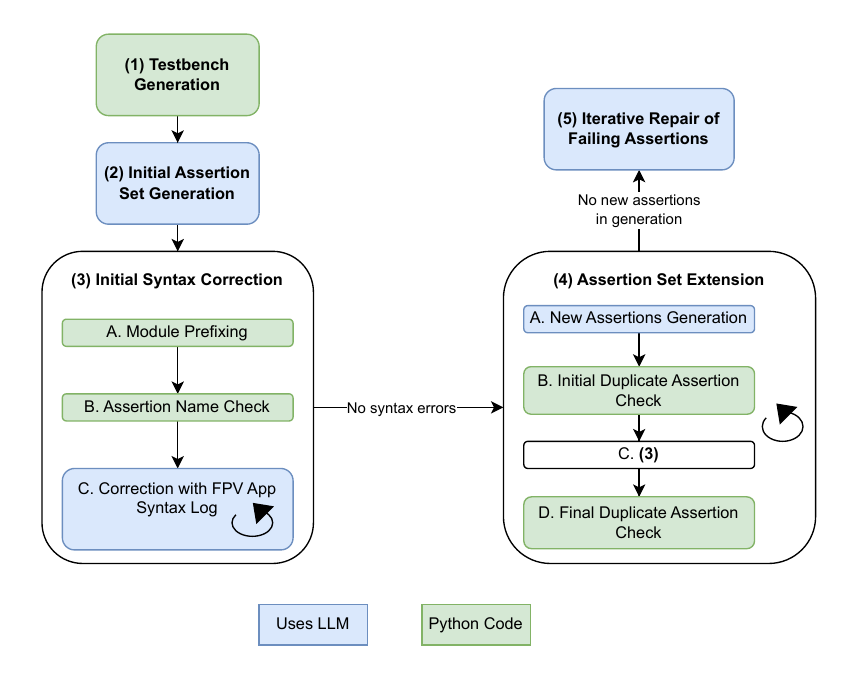}
\vspace{-8mm}
\caption{
\proj toolflow diagram showing the complete process from RTL input to final snapshot generation.
\textcolor{dgreen}{Green boxes} represent automated processes scripted in Python, \textcolor{blue}{blue boxes} represent processes involving LLMs.
}
\vspace{-5mm}
\label{fig:svapshot_toolflow}
\end{figure}
Our tool, \textbf{\proj}, is designed to be used in the rapid, iterative nature of AI-driven design. It addresses these limitations by introducing a framework focused on \textbf{\emph{creating a snapshot that captures the functionality of an RTL module as a comprehensive set of SystemVerilog Assertions (SVA) that serves as a regression suite for future modifications}.}
Unlike existing approaches that generate assertions from specifications, \proj leverages the goal of treating the current RTL as a golden reference to iteratively refine LLM-generated assertions and obtain a set of proven assertions with respect to the reference RTL. 
Specifications are often incomplete and expressed in diverse, non-textual forms such as waveforms or tables, which standard LLMs struggle to interpret unless these artifacts are transformed into text or handled by specialized models. Because natural language is inherently ambiguous, converting specs into a formal representation is critical to avoid misinterpretation. Meanwhile, RTL provides full visibility into the design’s microarchitecture, enabling LLMs to generate internal assertions that remain reusable as long as those parts of the design remain stable.
This snapshot methodology is particularly valuable for AI-driven design, where rapid changes are difficult to manually audit. \proj 's golden reference toolflow is orthogonal to prior works and could even adopt different assertion generation rules. The current version of \proj uses the rules from AutoSVA2~\cite{autosva2}.
\proj's generation based on reference RTL implementations is \textbf{\emph{also applicable to automatically creating formal testbenches for benchmarking of LLMs at hardware design}}, which is more exhaustive than simple functional equivalence comparison and provides more verification visibility than LEC (Logical Equivalence Checking), i.e. offers a direct view of the mismatching logic, with a concrete failing assertion, rather than a general output mismatch that requires debugging.
Several benchmarks have emerged to evaluate LLMs for RTL generation using coarser-grained metrics like functional equivalence or simulation-based testbenches.
These approaches would benefit from using \proj to curate assertion suites as \emph{property checklists} for exhaustively evaluating LLM-generated RTL.
RTLLM~\cite{Liu2024OpenLLMRTLCombined} provides an open benchmark suite comprising design descriptions, test cases, and golden RTL implementations to assess LLM capabilities in RTL generation.
VerilogEval~\cite{Liu2023VerilogEval} focuses on specific tasks like code completion and spec-to-RTL translation. 
TuRTLe~\cite{GarciaGasulla2025TuRTLe} offers a unified evaluation framework, integrating multiple existing benchmarks (including VerilogEval and RTLLM) to assess LLM performance across a spectrum of RTL generation tasks.
\proj is uniquely positioned to facilitate the benchmarking of LLM-generated RTL.
Instead of relying solely on simulation-based testbenches for benchmarking, which may miss corner-case bugs, prior work can leverage \proj to create comprehensive, high-quality, and formally verifiable assertion suites.
\proj introduces several \textbf{key innovations} that enable this snapshot methodology:
\begin{itemize}
    \item \textcircled{1} Implements an automated iterative repair loop that attempts to fix failing assertions using LLM-guided debugging, addressing a critical gap in existing tools that typically treat generated assertions as final outputs.
    \item \textcircled{2} Provides sophisticated duplicate detection and assertion set extension capabilities that ensure comprehensive coverage while avoiding redundancy through conjunctive normal form (CNF) analysis.
    \item \textcircled{3} Integrates these features through an automatic and modular pipeline that begins with automated formal testbench generation, proceeds through LLM-based assertion generation with systematic syntax correction, expands assertion sets through targeted prompting, and concludes with iterative semantic repair using formal verification feedback.
\end{itemize}
This end-to-end LLM-assisted automation, combined with robust error handling and systematic refinement, enables \proj to generate testbenches that serve both as verification artifacts and as benchmarking instruments for evaluating AI-generated RTL.
Figure~\ref{fig:svapshot_toolflow} illustrates the complete toolflow through a detailed diagram, which Algorithm~\ref{alg:svapshot_toolflow} further details with pseudocode.

\vspace{-2mm}
\section{The \proj Toolflow}\label{sec:svapshot_toolflow}
\proj is a modular framework that integrates LLM-based assertion generation with formal verification techniques.
It removes humans from the generation loop by treating the input RTL module as a golden reference.
By establishing trust in the RTL, we assume all proof results are correct, eliminating the notion of false positives (assertions that are passing but the RTL is wrong).
This assumption constitutes the cornerstone of \proj, which enables the semantic repair capability (Stage 5 of Algorithm \ref{alg:svapshot_toolflow}).
Section~\ref{sec:svapshot_toolflow_overview} details each toolflow stage, and Sections~\ref{sec:eval:snapshot} and \ref{sec:eval:benchmarking} present the use cases for which \proj's methodology introduces a novel solution.
\begin{figure}[t]
\centering
\begin{algorithm}[H]
\caption{\proj Toolflow}
\label{alg:svapshot_toolflow}
\begin{algorithmic}[1]
\REQUIRE RTL module $M$
\STATE \textbf{Stage 1:} Generate formal testbench environment for $M$
\STATE \textbf{Stage 2:} $A_{initial} \gets$ LLM.generateAssertions($M$, rules)
\STATE \textbf{Stage 3:} $A_{corrected} \gets$ correctSyntax($A_{initial}$)
\STATE \textbf{Stage 4:} Initialize property database with $A_{corrected}$
\REPEAT
    \STATE $A_{pot\_unique} \gets$ removeDupl($A_{new}$, prop\_db)
    \STATE $A_{pot\_unique\_corr} \gets$ correctSyntax($A_{pot\_unique}$)
    \STATE $A_{unique} \gets$ removeDupl($A_{pot\_unique\_corr}$, prop\_db)
    \STATE $A_{corrected} \gets A_{corrected} \cup$ $A_{unique}$
    \STATE Update property database (prop\_db)
\UNTIL{$A_{unique} = \emptyset$} \hfill $\triangleright$ All new assertions are duplicates
\STATE \textbf{Stage 5:} $results \gets$ formalVerify($A_{corrected}$, $M$)
\STATE $A_{proven} \gets \emptyset$, $A_{cex} \gets \emptyset$
\FOR{each assertion $a \in A_{corrected}$}
     \IF{$results[a] =$ PROVEN}
        \STATE $A_{proven} \gets A_{proven} \cup \{a\}$
    \ELSIF{$results[a] =$ CEX}
        \STATE $repaired \gets$ false, $attempts \gets 0$
        \WHILE{$\neg repaired$ \textbf{and} $attempts < 5$}
            \STATE $a_{fixed} \gets$ LLM.fixSVA($a$, $results[a].cex$, $M$)
            \STATE $result \gets$ formalVerify($a_{fixed}$, $M$)
            \IF{$result =$ PROVEN}
                \STATE $A_{proven} \gets A_{proven} \cup \{a_{fixed}\}$
                \STATE $repaired \gets$ true
            \ENDIF
            \STATE $attempts \gets attempts + 1$
        \ENDWHILE
        \IF{$\neg repaired$}
            \STATE $A_{cex} \gets A_{cex} \cup \{a\}$
        \ENDIF
    \ENDIF
\ENDFOR
\STATE \textbf{return} snapshot($A_{proven}$, $A_{cex}$)
\end{algorithmic}
\end{algorithm}
\vspace{-7mm}
\end{figure}
\subsection{Toolflow Overview}\label{sec:svapshot_toolflow_overview}
The \proj toolflow comprises five sequential stages:
\begin{enumerate}
    \item \textbf{Testbench Generation:} 
    Automated scripts generate the necessary formal testbench environment for the target RTL module.
    This includes directory structure, assertion files, binding files, and compilation scripts required for the Formal Property Verification (FPV) tool.
    
    \item \textbf{Initial Assertion Generation:} 
    The LLM generates an initial assertion set using the RTL module and curated SVA generation rules as input.
    The prompt combines the golden RTL with established SVA writing guidelines to enhance semantic relevance.
    
    \item \textbf{Syntax Correction:} 
    The framework corrects syntactic errors in the initial assertion set through a three-phase approach.
    Automated methods first address common errors: prefixing internal signals, enum values, parameters, and localparams with module names (\textbf{3.A} in Figure~\ref{fig:svapshot_toolflow}).
    
    The system then resolves assertion name collisions (\textbf{3.B} in Figure~\ref{fig:svapshot_toolflow}). These seldom happen, but we have observed instances of initial assertion sets with duplicated assertion names.
    Following compilation attempts with the FPV tool, remaining syntax errors trigger an LLM-based correction process that prompts the LLM with the RTL code, current assertions, and specific error messages (\textbf{3.C} in Figure~\ref{fig:svapshot_toolflow}).
    This iterative process continues until compilation succeeds. In the rare case that assertions prove very difficult to fix, a protection mechanism stops 
    the LLM-based syntax correction after 10 attempts. When this limit is reached, the tool progressively removes assertions from the set until a subset of syntactically correct assertions remains.
    \item \textbf{Assertion Set extension:} 
    Syntactically correct assertions are parsed and stored in a property database in conjunctive normal form (CNF) for reliable duplicate detection; CNF processing prevents wrongly considering assertions as different when they differ only in signal ordering (e.g., $a \land b$ and $b \land a$), or are logically equivalent (e.g., $a \implies b$ and $\lnot a \lor b$). The delay of every concurrent assertion is also stored for comparison. We remove the delay from the expression so that the parser only processes the implication, AND, OR, and NOT operands to do the CNF conversion. We define assertions as \emph{duplicated} when they are logically equivalent and have the same delay. If two assertions are different but have the same name, the tool adjusts the names automatically to prevent syntax errors.
    The LLM generates additional assertions using the RTL and existing properties as context.
    Automated duplicate detection precedes syntax correction to optimize computational efficiency, i.e., we avoid running the FPV tool on duplicate assertions.
    New assertions undergo syntax correction via Stage 3, followed by a final duplicate detection step.
    This assertion set extension process continues until the LLM produces only duplicate assertions.
    Thus, our de-duplication step serves as both a computationally-efficient filter and a reliable termination condition for the set extension process.
    \item \textbf{Iterative Semantic Repair:} 
    The semantic correction loop addresses failing assertions through LLM-based repair.
    Each iteration prompts the LLM with RTL code, one failing assertion, and previously attempted fixes.
    The framework limits repair attempts to 5 iterations per assertion to ensure termination.
    The toolflow concludes after attempting repairs for all failing assertions.
\end{enumerate}
\subsection{Use case: Snapshot-Based Verification of AI-Driven Design}\label{sec:eval:snapshot}
The primary use case for \proj is verifying the preservation of functionality in RTL designs undergoing AI-assisted modifications.
This application leverages formal assertions as regression tests to detect unintended behavioral changes.
\begin{itemize}
    \item \textbf{Establishing a Baseline:} 
    \proj generates a comprehensive set of SVA properties for an existing RTL module through the five-stage toolflow.
    After formal verification and iterative repair, these properties constitute an initial behavioral snapshot.
    This snapshot serves as a trusted baseline for subsequent design modifications.
    
    \item \textbf{Verifying AI-Generated Changes:} 
    Modified RTL undergoes verification against the established snapshot to detect behavioral regressions.
    End-to-end assertions involving only inputs and outputs remain applicable when the module interface is preserved.
    Internal assertions may require selective reuse based on the specific signals and implementation details affected by modifications.
    The key principle of \proj is that failing assertions identify potential issues introduced by changes, providing automated regression detection.
    \item \textbf{Incremental Snapshots:} 
    The snapshot methodology supports incremental updates as designs evolve and new functionalities undergo validation.
    Updated snapshots reflect the current trusted design state while maintaining historical verification artifacts.
\end{itemize}
This methodology enhances confidence in AI-assisted RTL development by providing automated formal verification of functional preservation across design iterations. This approach aligns well with test-driven development (TDD)~\cite{tdd_formal}.
\subsection{Use case: Benchmarking LLMs for RTL Generation}\label{sec:eval:benchmarking}
\proj enables fine-grained benchmarking of LLM capabilities in hardware design tasks beyond simple functional equivalence.
Functional equivalence (determined by tools like logic equivalence checkers~\cite{lec}) between reference and candidate designs provides only binary pass/fail metrics, which are insufficient to assess the quality and nuanced aspects of LLM output.
Although comprehensive benchmarking of LLM-generated RTL is beyond the scope of this paper, it provides a framework to enable such evaluation.
\textbf{Assertion-Based Scoring:} 
LLM-generated RTL can undergo evaluation against reference modules' snapshots to assess design quality.
The number of passing properties directly serves as a quality metric for RTL generation.
More sophisticated metrics could include weighted scoring based on assertion complexity (number of CNF clauses, number of signals, etc.).

\section{Experimental Methodology}\label{sec:eval:methodology}
\textbf{Module Selection:}
Table~\ref{tab:module_selection} lists the 14 RTL modules evaluated and provides a brief description for each of them. 
We evaluate \proj's capabilities using a diverse set of RTL modules spanning different complexity levels and design domains.
The module selection includes components from both internal company projects and open-source components from the Sargantana processor core~\cite{sargantana}, chosen to represent realistic hardware design scenarios.
\begin{table}[h]
\centering
\caption{RTL Modules Used for Evaluation.}
\label{tab:module_selection}
\small
\begin{tabular}{ll}
\toprule
\textbf{Module} & \textbf{Description} \\
\midrule
popcount         & Parameterizable pop-count/reduction\\
skipper\_counter & Configurable state counter\\
edge\_detector   & Signal edge detector\\
alu (Arithmetic Logic Unit) & Asynchronous ALU\\
dynamic\_target\_event\_counter & Parallel event counter\\
score\_board\_scalar & Scalar datapath scoreboard\\
branch\_unit &  Branch execution control unit\\
tlb (Transaction Lookaside Buffer) & Memory Management Unit TLB \\
ptw (Page Table Walker) & Memory Management Unit PTW\\
ptw\_arb & PTW request arbiter\\
pseudoLRU & Pseudo-LRU controller\\
div\_unit & Pipelined 32/64-bit division unit\\
div\_unit\_4bits & 4-bit division unit\\
mul\_unit & Pipelined 32/64-bit multiplication unit\\
\bottomrule
\end{tabular}
\end{table}
\textbf{Experimental Configuration:}
To sense what the impact of LLM sizes is on the \proj toolflow, we used the models in Table \ref{tab:llm-config}. The temperature value was chosen to balance assertion diversity with generation consistency and reproducibility, while the max token count was set based on the observation that each LLM task required around 4000 tokens. 
We ran the models using the NVIDIA NIM API~\cite{nvidianim}.
For formal property verification and coverage analysis, we utilized Cadence JasperGold v2021.12~\cite{jg_user}.
The repository \texttt{dac-2026-artifacts} contains the final assertion testbenches for \texttt{div\_unit}, \texttt{div\_unit\_4bits}, \texttt{mul\_unit}, \texttt{ptw}, \texttt{ptw\_arb}, and \texttt{pseudoLRU}, and the script that re-proves one experiment or the full set with JasperGold and with VC Formal.
\begin{table}[h]
\centering
\caption{LLMs used for experimentation.}
\small
\setlength{\tabcolsep}{1mm}%
\begin{tabular}{l c r}
\hline
\textbf{Model} & \textbf{Modules Evaluated} & \textbf{Configuration} \\
\hline
GPT-4.1-mini~\cite{gpt41mini}  & 14 & temp = 0.2, max\_token = 8192 \\
Llama3.1 405B\cite{llama3} & 6  & temp = 0.2, max\_token = 8192 \\
Llama3.3 70B~\cite{llama3}  & 6  & temp = 0.2, max\_token = 8192 \\
\hline
\end{tabular}
\label{tab:llm-config}
\end{table}

\noindent\textbf{Evaluation Metrics:}
Given that formal verification tools have been around for decades and have proved their usefulness at catching bugs~\cite{formal_book}, it would not add much value to the reader for us to prove that a formal testbench with full coverage would catch any bug that a code refactor may introduce.
Thus, we evaluate \proj effectiveness using checker coverage as the primary metric.
Checker coverage quantifies the portion of the design state space explored by the assertion set, providing a reliable indicator of assertion quality and completeness.
Higher checker coverage indicates more comprehensive verification of the RTL module's behavioral space.
Secondary metrics include initial assertion pass rates and semantic correction success rates.
These and other metrics enable analysis of both coverage progression and assertion set evolution throughout the flow, as well as assessment of the quality of generated assertions and the effectiveness of the repair process. The reported checker coverage results are calculated using only the proven assertions to establish the coverage baseline achieved by the snapshot.
Any remaining failing assertions, if corrected, would potentially contribute additional coverage to the design, especially since our methodology already detects and removes equivalent assertions.
\textbf{\emph{Measuring the coverage of the generated SVA evaluates the goodness of \proj at both use cases}} defined in Sections~\ref{sec:eval:snapshot} and \ref{sec:eval:benchmarking}, since the more thorough the formal testbench it creates, the more exhaustive is the snapshot at capturing future RTL deviations and the testbench at benchmarking nuanced RTL-generation capabilities.

\section{Experimental Results}\label{sec:results}
Section~\ref{sec:snapshot_coverage} presents comprehensive experimental results of applying \proj to a diverse set of RTL modules, ranging from 100 to 400 lines of code (LOC), including complex hierarchical designs with submodules. Section~\ref{sec:cex_correction} presents the CEX correction analysis.
Section~\ref{sec:robustness_and_time} presents an analysis of the runtime and cost of \proj, and lastly, a validation example is presented in Section~\ref{sec:validating_snapshot}.
Table~\ref{tab:svapshot_results} results demonstrate \proj's effectiveness across different module complexities and sources.
For the \texttt{tlb}, \texttt{ptw} and \texttt{div\_unit} modules, we generated assertions for their submodules as well, given their significantly larger size and complexity, providing a more comprehensive evaluation of \proj's capabilities on hierarchical designs.
Table~\ref{tab:svapshot_results} shows results for these modules with assertions for submodules, whereas for other modules without submodules (e.g., \texttt{skipper\_counter} and \texttt{dynamic\_target\_event\_counter}) we include only assertions for the top-level module.
Thus, further coverage could be achieved for these top-level modules by adding assertions for submodules, since the coverage tool reports coverage for the entire hierarchy.
\begin{table*}[t]
\centering
\caption{Experiment results. First section groups RTL modules evaluated with GPT4.1-mini; the rest cover the ones evaluated with 3 LLMs. Topology is either Sequential (S) or Combinational (C). Number of unique submodules (Sub). Initial (Init) results represent the results obtained after finishing Stage 3.}
\label{tab:svapshot_results}
\footnotesize
\setlength{\tabcolsep}{3pt}
\begin{tabular}{@{}llclcccccccc@{}}
\toprule
\textbf{RTL Module} & \textbf{LLM} & \textbf{Topology} & \textbf{Source} & \textbf{LOC} & \textbf{Sub} & \textbf{Init Assrt} & \textbf{Init Cov} & \textbf{Pass Before} & \textbf{Final Pass} & \textbf{Final Cov} \\
& & & & & \textbf{(\#)} & \textbf{(\#)} & \textbf{(\%)} & \textbf{Stage 5 (\#/Total)} & \textbf{(\#/Total)} & \textbf{(\%)} \\
\midrule
popcount                        & GPT 4.1 mini   & S    & Internal        & 139        & 2 & 5  & 0  & 0/59    & 3/59    & 93.9 \\
skipper\_counter                & GPT 4.1 mini   & S    & Internal        & 165        & 2 & 22 & 34.1 & 4/22    & 4/22    & 53.5 \\
edge\_detector                  & GPT 4.1 mini   & S    & Internal        & 24         & 0 & 5  & 73.3 & 19/19   & 19/19   & 73.3 \\
alu                             & GPT 4.1 mini   & C    & Internal        & 173        & 0 & 28 & 100  & 18/46   & 35/46   & 100 \\
dynamic\_target\_event\_counter & GPT 4.1 mini   & S    & Internal        & 230        & 1 & 14 & 62.8 & 6/14 & 6/14 & 62.8 \\
score\_board\_scalar            & GPT 4.1 mini   & S    & Sargantana Core & 96         & 0 & 22 & 55.7 & 11/22 & 12/22 & 86.0 \\
branch\_unit                    & GPT 4.1 mini   & C    & Sargantana Core & 140        & 0 & 14 & 14.4 & 1/14 & 12/14 & 42.0 \\
tlb (with submodules)        & GPT 4.1 mini   & S    & Sargantana Core & 372        & 1 & 8  & 23.7 & 33/43 & 37/43 & 39.3 \\
\quad pseudoLRU                 & GPT 4.1 mini   & S    & Sargantana Core & 137        & 0 & 8  & 68.7 & 22/35 & 29/35 & 70.7 \\
\midrule
ptw (with submodules)        & \textbf{GPT 4.1 mini}   & S    & Sargantana Core & 419        & 2 & 40 & 47.9 & 75/107  & 96/107  & \textbf{52.8} \\
\quad ptw\_arb                  & \textbf{GPT 4.1 mini}   & S    & Sargantana Core & 115        & 0 & 17 & 95.5 & 28/37   & 33/37   & \textbf{95.7} \\
\quad pseudoLRU                 & GPT 4.1 mini   & S    & Sargantana Core & 137        & 0 & 8  & 68.7 & 22/35 & 29/35 & 70.7 \\
ptw (with submodules)        & Llama 3.3 70b  & S    & Sargantana Core & 419        & 2 & 54 & 43.3 & 137/221 & 162/221 & 44.6 \\
\quad ptw\_arb                  & Llama 3.3 70b  & S    & Sargantana Core & 115        & 0 & 28 & 88.7 & 24/49   & 32/49   & 89.0 \\
\quad pseudoLRU                 & \textbf{Llama 3.3 70b}  & S    & Sargantana Core & 137        & 0 & 12 & 71.7 & 87/106  & 95/106  & \textbf{71.7} \\
ptw (with submodules)        & Llama 3.1 405b & S    & Sargantana Core & 419        & 2 & 5  & 28.8 & 68/116  & 86/116  & 43.4 \\
\quad ptw\_arb                  & Llama 3.1 405b & S    & Sargantana Core & 115        & 0 & 8  & 69.4 & 14/29   & 19/29   & 88.7 \\
\quad pseudoLRU                 & \textbf{Llama 3.1 405b} & S    & Sargantana Core & 137        & 0 & 11 & 68.7 & 21/45   & 32/45   & \textbf{71.7} \\
\midrule
div\_unit (with submodules)  & GPT 4.1 mini   & S    & Internal        & 241        & 1 & 17 & 84.2 & 74/128  & 96/128  & 85.6 \\
\quad div\_unit\_4bits          & GPT 4.1 mini   & C    & Internal        & 100        & 0 & 9  & 91.5 & 3/10    & 8/10    & 96.6 \\
div\_unit (with submodules)  & Llama 3.3 70b  & S    & Internal        & 241        & 1 & 60 & 76.0 & 129/170 & 132/170 & 83.9 \\
\quad div\_unit\_4bits          & \textbf{Llama 3.3 70b}  & C    & Internal        & 100        & 0 & 17 & 99.7 & 34/37   & 35/37   & \textbf{99.7} \\
div\_unit (with submodules)  & \textbf{Llama 3.1 405b} & S    & Internal        & 241        & 1 & 36 & 74.5 & 31/65   & 51/65   & \textbf{89.5 }\\
\quad div\_unit\_4bits          & \textbf{Llama 3.1 405b} & C    & Internal        & 100        & 0 & 16 & 91.6 & 20/23   & 21/23   & \textbf{99.7} \\
\midrule
mul\_unit                       & \textbf{GPT 4.1 mini}   & S    & Internal        &    294        & 0 & 40 & 92.6 & 51/115  & 81/115  & \textbf{98.9} \\
mul\_unit                       & Llama 3.3 70b  & S    & Internal        & 294        & 0 & 29 & 0.7  & 4/45   & 10/45   & 1.76 \\
mul\_unit                       & Llama 3.1 405b & S    & Internal        & 294        & 0 & 10 & 1.5  & 24/67   & 37/67   & 70.3 \\
\bottomrule
\end{tabular}
\vspace{-3mm}
\end{table*}
\vspace{-2mm}
\subsection{Snapshot Coverage Across Modules}\label{sec:snapshot_coverage}
The experimental results demonstrate several key findings across multiple dimensions of analysis.
\textbf{Effect of LLM choice:}
The first section of Table~\ref{tab:svapshot_results} shows modules evaluated only with GPT 4.1 mini, while the latter three sections show modules evaluated with several LLMs (and bold the highest coverage for each module).
We notice that while GPT 4.1 mini performs well overall, for some modules, the Llama models surpass the achieved coverage.
The 405-billion parameter Llama version is better than or equal to the 70-billion one, and results are vastly better for \texttt{mul\_unit}.
We anticipate that future, more capable AI models (especially those fine-tuned on SystemVerilog) would yield even higher coverage, with the same tooflow.
\textbf{Coverage Improvement Analysis:}
\proj achieved significant coverage improvement from initial SVA generation (Stage 2 in Figure~\ref{fig:svapshot_toolflow}) to the end of set extension (Stage 4) across the majority of evaluated modules.
The \texttt{alu} module achieves 100\% initial coverage, an exceptional result, due to its simple functional specifications, which are easily translated into SVA properties.
The most substantial improvement occurred in the \texttt{popcount} module, where coverage increased from up to 93.9\%. This module only has one output signal, and the three correct assertions targeted it, thus increasing the coverage significantly.
Modules that exhibited modest coverage increases typically experienced limited assertion set extension during the extension phase, rather than extension without corresponding coverage gains, as occurred, for example, in the \texttt{div\_unit\_4bits} module.
The \texttt{edge\_detector} module maintained its initial 73.3\% coverage, as the generated assertions were already comprehensive and no additional meaningful assertions were discovered during extension.
\textbf{RTL Topology Impact:}
The results reveal distinct patterns based on RTL topology and functional domain.
Units \texttt{alu} and \texttt{div\_unit} exhibited high initial coverage and strong assertion correction rates, though not necessarily the highest coverage increases due to already comprehensive initial coverage.
Control modules like \texttt{ptw} demonstrated greater resistance to semantic correction, likely due to their complex behavioral specifications.
\textbf{Assertion Generation Quality:}
The initial pass rates of generated assertions were generally high across most modules, indicating that LLMs can produce syntactically and semantically reasonable assertions with minimal repair.
Modules with lower initial pass rates typically involve complex control logic or timing-sensitive behaviors that are inherently more difficult to specify correctly.
The semantic correction phase successfully improved assertion pass rates in most cases, with notable exceptions in modules requiring domain-specific assumptions about input constraints.
\textbf{Scalability Considerations:}
The analysis of the \texttt{tlb} and \texttt{ptw} modules reveals the challenges associated with scaling to larger module sizes.
\proj addresses entire modules without partitioning, thus demonstrating its ability to achieve meaningful coverage even on complex designs when applied holistically.
Moreover, since traditional FPV approaches for modules of this size may already employ divide-and-conquer strategies due to state space explosion concerns, further modularization of the design could enable \proj to achieve better results as observed with smaller modules.
\vspace{-2mm}
\subsection{Counterexample (CEX) Correction}\label{sec:cex_correction}
Converting CEXs into passing assertions boosts the coverage of the SVA testbench created by \proj.
Our methodology performs several LLM-driven attempts to fix failing (CEX) assertions (Stage 5 in Algorithm~\ref{alg:svapshot_toolflow}).
However, this process may not address all assertion CEXs, since assertions may have semantic errors, or assumptions could be missing for semantically correct assertions.
Thus, after using \proj, an engineer could carefully review whether assumptions are missing for module inputs that would lead to spurious CEXs, as well as revise failing assertions for semantic errors.
Next, we analyze whether there are trivial assumption-fixable CEXs and the effectiveness of LLM-driven corrections.
\textbf{Assumptions can fix CEXs and increase coverage:}\label{sec:fix1}
The FPV tool revealed that some assertions are semantically correct but fail (i.e., lead to CEX) due to underconstrained inputs.
In the scoreboard module, the failing assertion \texttt{as\_\_OcupDivUnitQ1ShiftRightNoSet} was corrected by adding an assumption constraining the flush signal to zero, confirming the hypothesis that input assumptions can resolve some CEXs reported by the FPV tool.
This finding suggests that \proj's current limitation in generating input assumptions leads to false negative results that could be addressed through assumption synthesis.
However, as the literature~\cite{formal_book} tells us, prematurely adding assumptions may lead to false positives, so we decided not to automatically add assumptions in our methodology and leave that to the engineer.%
\textbf{Semantic Correction Attempt Analysis:}\label{sec:fix2}
After extracting execution data for the final phase of the flow, we measured the effectiveness of attempting to correct failing assertions (CEX).
The majority of the assertions that the LLM corrected (went from CEX to Proof) were done so on the first attempt.
For modules evaluated on all LLMs, GPT 4.1 mini corrected 27\% of the assertions on the first attempt, and the total percentage of corrected assertions in 5 attempts was 43.3\%; for Llama 3.3 70b 17.3\% and 25.8\%; and for Llama 3.1 405b 15.6\% and 43.8\%.
We could potentially reduce the attempt count significantly from five attempts (and expect most of the LLM-fixable properties to be corrected) to reduce the LLM cost (albeit low as shown in Section~\ref{sec:robustness_and_time}), LLM latency, and FPV runtime, since this phase takes the longest, and is entirely dependent on the LLM and the FPV tool. 
\vspace{-2mm}
\subsection{\proj Runtime and Cost}\label{sec:robustness_and_time}
Dissecting the time that each of the stages in the \proj toolflow takes (detailed in Algorithm~\ref{alg:svapshot_toolflow}), we found that generating the initial testbench (Stage 1 and 2) takes the order of a few seconds, being the most residual execution time. 
Initial syntax correction of the initial assertion set (Stage 3) required on average 2 minutes (3.3\% of the total time)
The set extension (Stage 4) consumed 9 minutes on average (19.3\% of the total).
Lastly, Semantic correction (Stage 5) with 5 attempts required 55 minutes (77.4\% of the total), due to the LLM call latency and FPV tool runtime.
This is why the reduction of the attempt limit we give the LLM in Stage 5 would be a way to reduce the execution time and obtain faster snapshots.
Further dissection of Stages 3-5 showed that the FPV tool usage represented 69.4\% of the total execution time on average.
Despite these high execution times, we evaluated the cost of each experiment based on the input and output tokens used and prices from inference providers for each of the models, and obtained that 
the average cost for a snapshot was 0.5€, which again could be reduced if the attempt limit in Stage 5 were tuned down.
\subsection{Validating The Snapshot Methodology}\label{sec:validating_snapshot}
\textbf{Validation Through Bug Injection:}
To evaluate the practical utility of generated assertions, we manually introduced a representative bug in the scoreboard module.
The bug involved incorrectly setting \texttt{inst\_d[2]} instead of \texttt{inst\_d[1]} when a 64-bit multiply instruction is issued, creating an off-by-one timing error.
The assertion \texttt{as\_\_SetMul64SetsInstD1} (Listing~\ref{lst:scoreboard_sva}) successfully detected this bug, demonstrating that \proj generates assertions capable of capturing meaningful design intent and detecting realistic implementation errors.
\begin{lstlisting}[language=Verilog, label=lst:scoreboard_sva, caption=Snapshot assertion for the scoreboard module.]
as__SetMul64SetsInstD1: assert property((set_mul_64_i) |-> (scoreboard_sargantana.inst_d[1] == 1'b1));
\end{lstlisting}
\textbf{Validation Through LLM-based Refactoring:}
We performed an experiment of LLM-based refactor of the \texttt{ptw\_arb} module using Cursor.
We asked the code agent (claude-4-sonnet) to refactor the original \texttt{ptw\_arb} module, keeping the module interface signals intact.
After refactoring its finite-state machine (FSM) and running the snapshot from the original module through the FPV tool, we saw a CEX for an assertion that passed originally (Listing~\ref{lst:ptw_arb_cex}).
\par\begin{minipage}[b]{\columnwidth}
\begin{lstlisting}[language=Verilog, label=lst:ptw_arb_cex, caption=Failing snapshot assertion after LLM refactor of RTL.]
as__IdleStateOnlyDtlbReqOutput:assert property(
    (ptw_arb.current_state_q==ptw_arb.IDLE &&
     dtlb_ptw_i.req.valid && !itlb_ptw_i.req.valid &&
     !ptw_arb.itlb_req_pending_q) |->
    (tlb_ptw_o==dtlb_ptw_i));
\end{lstlisting}
\end{minipage}\par
\vspace{-2mm}
This assertion captures the behavior that when the PTW arbiter is in \texttt{IDLE} state and only the DTLB has a valid request (no ITLB request), the output should directly connect the DTLB communication interface.
This behavior breaks in the version edited by AI since the refactored version splits the \texttt{next\_state} signals from the output signals into two blocks and then confuses the state for which \texttt{tlb\_ptw\_comm\_o} was supposed to change, from \texttt{IDLE} to \texttt{SERVING\_DTLB}.
This example demonstrates the value of having the snapshot before refactoring or modifications, as well as the ability for the snapshot to detect bugs in the refactored version.

\vspace{-1.5mm}
\section{Discussion and Conclusion}\label{sec:conclusion}
This paper introduces \proj, a systematic framework for automated SVA generation that addresses critical verification challenges in AI-driven RTL design workflows.
The increasing reliance on LLMs for RTL code generation and modification necessitates robust mechanisms for ensuring functional correctness and regression testing.
Although SVA generation has been explored previously, our snapshot-based approach enables a systematic refinement of assertions to conform to an RTL implementation. 
\proj provides a framework for capturing design functionality of an RTL module as formal assertions, enabling two use cases: (1) establishing an implementation baseline for regression testing of subsequent modifications, and (2)  generating testbenches for modules used to benchmark LLM-aided RTL design.
Our methodology accomplishes this by implementing an iterative repair approach that systematically addresses both syntactic and semantic errors in LLM-generated assertions through automated correction loops, significantly improving assertion quality over single-pass generation methods.
Overall, \proj's toolflow combines assertion generation, duplicate detection, assertion set extension, and formal verification in a cohesive framework, reducing manual intervention.
Our experimental evaluation across diverse RTL modules demonstrates the framework's effectiveness in achieving high coverage across a set of 14 diverse, complex modules with up to 400 lines of SystemVerilog.
The semantic correction phase successfully improved assertion pass rates in most evaluated designs, while the assertion set extension mechanism generated comprehensive property sets without logical redundancy.
The validation through deliberate bug injection confirms that \proj generates assertions capable of detecting realistic implementation errors, both human- and LLM-generated.
Moreover, the experiments reveal interesting aspects of LLM-based assertion generation.
Arithmetic and datapath modules consistently achieved higher initial coverage and correction rates compared to complex control logic implementations.
When comparing the testbenches generated by various LLMs, we observed higher coverage with newer (GPT 4.1 mini) or larger (Llama 405b) LLMs, which suggests that future improvements in LLM capabilities could further enhance coverage results of the \proj methodology.
However, we acknowledge some \textbf{limitations} of our methodology:
(1)~\proj does not verify the correctness of the input RTL, but rather aims to verify that future modifications do not break the current functionality. 
(2)~The semantic repair process, while effective in most cases, does not address CEXs requiring domain-specific input assumptions.
Future research directions include extending the framework to generate adequate assumptions on RTL module inputs (to avoid spurious CEXs), as well as developing more sophisticated metrics for assertion relevance when using assertions as weighted scores for LLM benchmarking.
The application of \proj to comprehensive LLM benchmarking could be a paper on its own, to standardize evaluation methodologies for AI-generated RTL designs.
While challenges remain in handling complex control logic and assumption generation, the demonstrated effectiveness and inexpensive execution across a diverse set of RTL modules suggest that this work represents a significant step toward trustworthy AI-assisted hardware verification, providing both the theoretical framework and practical tools necessary to maintain design integrity in an era of increasingly automated hardware development.

\newpage
\bibliographystyle{ACM-Reference-Format}
\bibliography{references.bib}

@book{formal_book,
  title={Formal verification: an essential toolkit for modern VLSI design},
  author={Seligman, Erik and Schubert, Tom and Kumar, MV Achutha Kiran},
  year={2015},
  publisher={Morgan Kaufmann}
}

@article{ilang,
  title={Instruction-Level Abstraction {(ILA)}: A Uniform Specification for System-on-Chip Verification},
  author={Huang, Bo-Yuan and Zhang, Hongce and Subramanyan, Pramod and Vizel, Yakir and Gupta, Aarti and Malik, Sharad},
  journal={ACM Transactions on Design Automation of Electronic Systems (TODAES)},
  volume={24},
  number={1},
  pages={1--24},
  year={2018},
  publisher={ACM New York, NY, USA}
}

@misc{jg_user,
  title={JasperGold Apps User’s Guide},
  author={{Cadence Design Systems Inc.}},
  year={2015}
}

@misc{sargantana,
  title={Sargantana},
  author={{Barcelona Supercomputing Center}},
  year={2025},
  note={\url{https://github.com/bsc-loca/sargantana}}
}

@book{lec,
  title={Formal equivalence checking and design debugging},
  author={Huang, Shi-Yu and Cheng, Kwang-Ting Tim},
  volume={12},
  year={2012},
  publisher={Springer Science \& Business Media}
}

@inproceedings{tdd_formal,
  title={Formal for Designers},
  author={Buckingham, Jon},
  booktitle={Agile Test Driven Dev. for ASIC},
  year={2016},
  doi={https://www.testandverification.com/wp-content/uploads/2016/VerificationFutures/9_HPE_JohnBuckingham_(ASICPresentation).pdf}
}

@INPROCEEDINGS{autosva,
  author={Orenes-Vera, Marcelo and Manocha, Aninda and Wentzlaff, David and Martonosi, Margaret},
  booktitle={ACM/IEEE Design Automation Conference (DAC)}, 
  title={{AutoSVA: Democratizing Formal Verification of RTL Module Interactions}}, 
  year={2021},
  volume={},
  number={},
  pages={535-540},
  doi={10.1109/DAC18074.2021.9586118}}

@misc{autosva2,
      title={Using LLMs to Facilitate Formal Verification of RTL}, 
      author={Marcelo Orenes-Vera and Margaret Martonosi and David Wentzlaff},
      year={2023},
      eprint={2309.09437},
      archivePrefix={arXiv},
      primaryClass={cs.AR},
      url={https://arxiv.org/abs/2309.09437}, 
}

@article{goldmine2013,
  title={Mining hardware assertions with guidance from static analysis},
  author={Li, Wenchao and Dworkin, Lee and Seshia, Sanjit A.},
  journal={IEEE Transactions on Computer-Aided Design of Integrated Circuits and Systems},
  volume={32},
  number={6},
  pages={952--965},
  year={2013},
  publisher={IEEE}
}

@article{harm2022,
  title={HARM: A Hint-Based Assertion Miner},
  author={Germiniani, Samuele and Pravadelli, Graziano},
  journal={IEEE Transactions on Computer-Aided Design of Integrated Circuits and Systems},
  volume={41},
  number={11},
  pages={4277--4288},
  year={2022},
  publisher={IEEE}
}

@inproceedings{assertllm2025,
  title={AssertLLM: Generating Hardware Verification Assertions from Design Specifications via Multi-LLMs},
  author={Yan, Zhiyuan and Fang, Wenji and Li, Mengming and Li, Min and Liu, Shang and Xie, Zhiyao and Zhang, Hongce},
  booktitle={30th Asia and South Pacific Design Automation Conference (ASPDAC '25)},
  year={2025}
}

@article{assertionforge2025,
  title={AssertionForge: Enhancing Formal Verification Assertion Generation with Structured Representation of Specifications and RTL},
  author={Bai, Yunsheng and Hamad, Ghaith Bany and Suhaib, Syed and Ren, Haoxing},
  journal={arXiv preprint arXiv:2503.19174},
  year={2025}
}

@article{rtlcoder2024,
  title={RTLCoder: Fully Open-Source and Efficient LLM-Assisted RTL Code Generation},
  author={Liu, Shang and Fang, Wenji and Lu, Yao and Wang, Jing and Zhang, Qijun and Zhang, Hongce and Xie, Zhiyao},
  journal={arXiv preprint arXiv:2312.08617},
  year={2024}
}

@article{aivril2024,
  title={AIvril: AI-Driven RTL Generation with Verification In-The-Loop},
  author={Islam, Mubashir ul and Sami, Humza and Gaillardon, Pierre-Emmanuel and Tenace, Valerio},
  journal={arXiv preprint arXiv:2409.11411},
  year={2024}
}

@article{rtlpp2025,
  title={RTL++: Graph-enhanced LLM for RTL Code Generation},
  author={Akyash, Mohammad and Azar, Kimia and Kamali, Hadi},
  journal={arXiv preprint arXiv:2505.13479},
  year={2025}
}

@inproceedings{Danese2017ATeam,
title = {{A-TEAM: Automatic template-based assertion miner}},
author = {Alessandro Danese and Nicol{`o} Dalla Riva and Graziano Pravadelli},
booktitle = {Proc. DAC},
pages = {37:1--37:6},
year = {2017},
doi = {10.1145/3061639.3062206}
}

@article{Menon2025Spec2Assertion,
  author    = {Menon, Anand and Pal, Debjit and Nandal, Deeksha and Pulavarthi, Vaishnavi and Dan, Soham and Kundu, Shamik and Srivastava, Amisha},
  title     = {Spec2Assertion: Generating SystemVerilog Assertions from Design Specifications using Large Language Models with Chain-of-Thought Prompting and Progressive Regularization},
  journal   = {arXiv preprint arXiv:2505.07995v2},
  year      = {2025},
  month     = {May},
  eprint    = {2505.07995},
  archivePrefix = {arXiv},
  primaryClass = {cs.AI},
  url       = {[https://arxiv.org/abs/2505.07995](https://arxiv.org/abs/2505.07995)}
}

@inproceedings{Liu2024OpenLLMRTLCombined,
  author    = {Liu, Shang and Lu, Yao and Fang, Wenji and Li, Mengming and Xie, Zhiyao},
  title     = {OpenLLM-RTL: Open Dataset and Benchmark for LLM-Aided Design RTL Generation and Verification},
  booktitle = {Proceedings of the ACM/IEEE International Conference on Computer-Aided Design (ICCAD ’24)},
  year      = {2024},
  month     = {Oct},
  note      = {Preprint arXiv:2503.15112},
  url       = {[https://arxiv.org/abs/2503.15112](https://arxiv.org/abs/2503.15112)}
}

@article{GarciaGasulla2025TuRTLe,
  author    = {Garcia-Gasulla, Dario and Kestor, Gokcen and Parisi, Emanuele and Albert{\'\i}-Binimelis, Miquel and Gutierrez, Cristian and Ghorab, Razine Moundir and Montenegro, Orlando and Homs, Bernat and Moret{\'o}, Miquel},
  title     = {TuRTLe: A Unified Evaluation of LLMs for RTL Generation},
  journal   = {arXiv preprint arXiv:2504.01986v1},
  year      = {2025},
  month     = {Mar},
  eprint    = {2504.01986},
  archivePrefix = {arXiv},
  primaryClass = {cs.AR},
  url       = {[https://arxiv.org/abs/2504.01986](https://arxiv.org/abs/2504.01986)}
}

@article{Liu2023VerilogEval,
  author    = {Liu, Shang and Yan, Zhiyuan and Fang, Wenji and Lu, Yao and Li, Mengming and Zhang, Hongce and Xie, Zhiyao},
  title     = {VerilogEval: A Large-Scale Dataset for Verilog Code Generation and Understanding},
  journal   = {arXiv preprint arXiv:2310.09231},
  year      = {2023},
  month     = {Oct},
  eprint    = {2310.09231},
  archivePrefix = {arXiv},
  primaryClass = {cs.SE},
  url       = {[https://arxiv.org/abs/2310.09231](https://arxiv.org/abs/2310.09231)}
}

@misc{gpt41mini,
  title={GPT-4.1-mini},
  author={OpenAI},
  year={2025},
  note={\url{https://platform.openai.com/docs/models/gpt-4.1-mini}}
}

@misc{llama3,
  title={Llama 3.1},
  author={Meta},
  year={2025},
  note={\url{https://ai.meta.com/blog/meta-llama-3-1/}}
}

@misc{nvidianim,
  title={NVIDIA NIM},
  author={NVIDIA},
  year={2025},
  note={\url{https://build.nvidia.com/explore/discover}}
}
\end{document}